\documentclass[aps,prl,twocolumn,showpacs,superscriptaddress]{revtex4-1}

\usepackage[final]{graphicx}
\usepackage{amsmath}
\usepackage{verbatim}
\usepackage{color}
\usepackage{ulem}
\usepackage[colorlinks,
linkcolor=blue,
citecolor=blue,
urlcolor=blue]{hyperref}

\begin{document}

\preprint{}

\title{Suppression of Stimulated Raman Scattering and Hot Electrons Generation due to Langmuir Decay Instability Cascade and Langmuir Collapse}

\author{Q. S. Feng} 
\affiliation{HEDPS, Center for
	Applied Physics and Technology, Peking University, Beijing 100871, China}

\author{C. Y. Zheng} \email{zheng\_chunyang@iapcm.ac.cn}

\affiliation{HEDPS, Center for
	Applied Physics and Technology, Peking University, Beijing 100871, China}
\affiliation{Institute of Applied Physics and Computational
	Mathematics, Beijing, 100094, China}
\affiliation{Collaborative Innovation Center of IFSA (CICIFSA) , Shanghai Jiao Tong University, Shanghai, 200240, China}

\author{Z. J. Liu} 
\affiliation{HEDPS, Center for
	Applied Physics and Technology, Peking University, Beijing 100871, China}
\affiliation{Institute of Applied Physics and Computational
	Mathematics, Beijing, 100094, China}

\author{L. H. Cao} 
\affiliation{HEDPS, Center for
	Applied Physics and Technology, Peking University, Beijing 100871, China}
\affiliation{Institute of Applied Physics and Computational
	Mathematics, Beijing, 100094, China}
\affiliation{Collaborative Innovation Center of IFSA (CICIFSA) , Shanghai Jiao Tong University, Shanghai, 200240, China}

\author{Q. Wang}
\affiliation{HEDPS, Center for
	Applied Physics and Technology, Peking University, Beijing 100871, China}
\author{C. Z. Xiao}
\affiliation{School of Physics and Electronics, Hunan University, Changsha 410082, China}
\affiliation{HEDPS, Center for
	Applied Physics and Technology, Peking University, Beijing 100871, China}
\author{X. T. He} \email{xthe@iapcm.ac.cn}
\affiliation{HEDPS, Center for
	Applied Physics and Technology, Peking University, Beijing 100871, China}
\affiliation{Institute of Applied Physics and Computational
	Mathematics, Beijing, 100094, China}
\affiliation{Collaborative Innovation Center of IFSA (CICIFSA) , Shanghai Jiao Tong University, Shanghai, 200240, China}


\date{\today}

\begin{abstract}
Backward stimulated Raman scattering (BSRS) with Langmuir decay instability (LDI) and Langmuir collapse has been researched by Vlasov simulation for the first time. The decay productions of LDI cascade and their evolution with time is clearly demonstrated, which occurs simultaneously with Langmuir collapse. The BSRS reflectivity will be decreased largely through LDI cascade and Langmuir collapse. In CH plasmas, when $T_i/T_e=1/3$, the Landau damping of the slow ion-acoustic wave (IAW) is lower than that in H plasmas. Therefore, the BSRS can be further suppressed through LDI cascade by the way of controlling the species of plasmas and ions ratio. These results give an effective mechanism to suppress the BSRS and hot electrons generation.
	
\end{abstract}

\pacs{52.38.Bv, 52.35.Fp, 52.35.Mw, 52.35.Sb}

\maketitle


Backward stimulated Raman scattering (BSRS)\cite{Troccoli_2005nature,Vieira_2016NC,Yampolsky_2014NP,Depierreux_2014NC}, i.e., the incident electromagnetic (EM) wave decays into a Langmuir wave (LW) and a inverse scattered EM wave, is detrimental in the inertial confinement fusion (ICF) \cite{He_2016POP, Glenzer_2007Nature,Glenzer_2010Science} experiments. Because BSRS will lead to a net energy loss of the incident laser beams and may affect the irradiation symmetry, in addition, the LWs will generate fast electrons which is able to preheat the fusion fuel. The suppression of BSRS is an important component of the laser-driven ICF research. A possible saturation mechanism for BSRS is the Langmuir decay instability (LDI, the process by which the primary LW decays into an ion acoustic wave (IAW) and a secondary LW) \cite{Kirkwood_1996PRL,Depierreux_2000PRL,Montgomery_2001PRL,Depierreux_2002PRL,Fouquet_2008PRL} or the Langmuir collapse \cite{Russell_1999POP}. Especially, Fouquet \cite{Fouquet_2008PRL} researched the effect of LDI on BSRS in an inhomogeneous plasma, and found that LDI can suppress the gradient stabilization, thus leading to a significantly increased BSRS reflectivity. However, in homogeneous plasmas, the results were opposite. Kirkwood \cite{Kirkwood_1996PRL} found that the BSRS reflectivity in a homogenous plasma depended directly on IAW damping and increased with the IAW damping. In this Letter, our view is that with the IAW damping increasing, the LDI will be suppressed, thus less BSRS LW energy will transfer to the IAW and decay LW, as a result, the BSRS reflectivity will increase in homogeneous plasmas. We have also demonstrated the clear physical pictures of LDI cascade and its decay productions through Vlasov simulation, which is an important saturation mechanism of BSRS.

Corresponding experiments carried out by Depierreux et al. \cite{Depierreux_2000PRL} reported the observation of IAWs resonantly produced by LDI for the first time, thus they thought the presence of LDI linked to BSRS was fully confirmed. However, Montgomery \cite{Montgomery_2001PRL} gave a comment on the work in Ref. \cite{Depierreux_2000PRL} and he thought the IAW spectra did not conclusively support the case for LDI cascade. In Montgomery's opinion, the IAW spectra could be the result of either strong turbulence due to Langmuir collapse \cite{Bezzerides_1993PRL,Russell_1999POP} or multiple LDI cascades in an inhomogeneous plasma. In this Letter, we will give a clear demonstration of the LDI cascade productions and the LDI cascade evolution with time for the first time. And at the same time, the Langmuir collapse also occurs simultaneously in our simulation parameter $k_{L1}\lambda_{De}=0.18$ ($k_{L1}$ is the wave number of BSRS LW and $\lambda_{De}$ is the electrons Debye length), which is different from Ref. \cite{Depierreux_2002PRL} where $k_{L1}\lambda_{De}>0.2$ and the Langmuir collapse did not occur in that parameter region.

\begin{figure}[!tp]
 	\includegraphics[width=1\columnwidth]{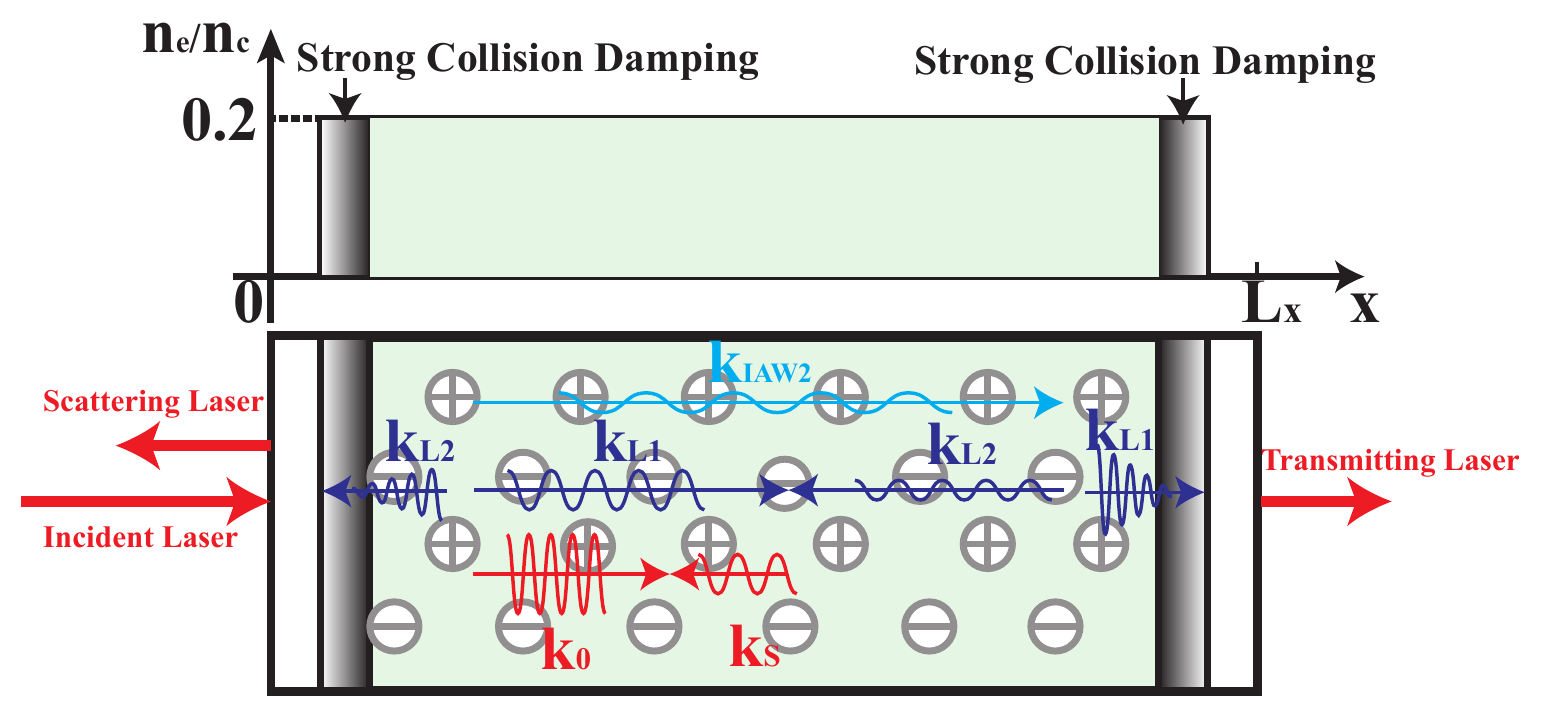}
 	
 	\caption{\label{Fig:Schematic}(Color online) The schematic of the BSRS and LDI process. The strong collision damping layer is to absorb the LWs and IAWs, and prevent the electrostatic waves from reflecting.}
 \end{figure}

A model is demonstrated to show the BSRS and LDI process. As shown in Fig. \ref{Fig:Schematic}, a strong collision damping layer is added to the two sides of the plasmas boundaries, as a result, the plasmas waves such as Langmuir waves and ion-acoustic waves will be damped and nearly not be reflected in the boundaries. An one dimension in space and three dimensions in velocity (1D3V) Vlasov-Maxwell code \cite{Liu_2009POP,Liu_2009POP_1} is used to simulate the BSRS and LDI process in different cases of species. We have taken the H plasmas and CH (1:1) plasmas as typical examples. To compare the BSRS process without LDI process, we have also given the example of the fixed background ions. The electrons density is $n_e=0.2n_c$, $n_c$ is the critical density for the incident laser. In these cases, the density is as large as $0.2n_c$ so that the BSRS rescatter\cite{Winjum_2013PRL} doesn't occur. And the electrons temperature is $T_e=2.5keV$, so the wave number of BSRS LW (denoted as LW1) is $k_{L1}=1.18c/\omega_0=0.18\lambda_{De}^{-1}$ ($c$ and $\omega_0$ is the vacuum velocity and frequency of the incident laser, $\lambda_{De}$ is the electrons Debye length). The ions temperature $T_i$ is assumed to be the same and $T_i/T_e=1/3$. The linearly polarized laser intensity in our simulation is $I=3\times10^{15}W/cm^2$ with wavelength $\lambda_0=0.351\mu m$. The spatial scale is $[0, L_x]$ ($L_x=500c/\omega_0$ with $2\times5\%L_x$ vacuum layers and $2\times5\%L_x$ collision layers in the two sides of plasmas boundaries) discretized with $N_x=5000$ grid points and space step $dx=0.1c/\omega_0$. The total simulation time is $t_{end}=5\times10^{4}\omega_0^{-1}$ discretized with $N_t=5\times10^5$ and time step $dt=0.1\omega_0^{-1}$. And the velocity scale is discretized with $N_v=512$ grid points. Fig. \ref{Fig:Schematic} demonstrates the schematic of BSRS and LDI process. Firstly, the incident electromagnetic (EM) wave decays into a scattered EM wave and a Langmuir wave (LW) through BSRS. The matching conditions for BSRS is: $\omega_0=\omega_s+\omega_{L1}$ and $\vec{k}_0=\vec{k}_s+\vec{k}_{L1}$, where $\omega_i$ and $\vec{k}_i$ are the frequencies and wave numbers of the incident laser ($i=0$), scattering laser ($i=s$) and BSRS Langmuir wave ($i=L1$). Secondly, the LW (denoted as $L1$) generated by BSRS will decay to a secondary Langmuir wave ($L2$) and an ion-acoustic wave ($IAW2$) if the density fluctuation of pump LW satisfies the threshold for LDI\cite{Karttunen_1981PRA}:
\begin{equation}
(\frac{\delta n}{n})_{LDI}=2k_{L1}\lambda_{De}\sqrt{(\frac{\nu_{ia2}}{\omega_{ia2}})(\frac{\nu_{L2}}{\omega_{pe}})}
\end{equation}
where $k_{L1}$ is the wave number of the LW generated by BSRS, $\nu_{L2}$ is the linear damping of the secondary LW generated by LDI, $\nu_{ia2}$ and $\omega_{ia2}$ is the linear damping rate and the frequency of IAW generated by LDI. $\omega_{pe}=\sqrt{4\pi n_ee^2/m_e}$ is the plasmas frequency of electrons. Only Landau damping of LW and IAW is considered. Similarly, the matching conditions for LDI is: $\omega_{L1}=\omega_{L2}+\omega_{IAW2}$ and $\vec{k}_{L1}=\vec{k}_{L2}+\vec{k}_{IAW2}$. And the wave numbers of the secondary LW and IAW generated by LDI are:
$|\vec{k}_{L2}|=|\vec{k}_{L1}|-\Delta k$,
$|\vec{k}_{IAW2}|=2|\vec{k}_{L1}|-\Delta k$,
where $\Delta k\ll |\vec{k}_{L1}|$. And the wave vectors of the LWs ($\vec{k}_{Ln}$) and IAWs ($\vec{k}_{IAWn}$) generated in step $n-1$ of the LDI cascade are:
$|\vec{k}_{Ln}|=|\vec{k}_{L1}|-(n-1)\Delta k$,
$|\vec{k}_{IAWn}|=2|\vec{k}_{L1}|-(2n-3)\Delta k$.

Through the dispersion relation of the LW and the IAW:
$\omega_L^2=\omega_{pe}^2+3k_L^2v_{te}^2$,
$\omega_{s}=k_{s}c_s$,
where subscript $L$ denotes the Langmuir waves of LDI, and subscript $s$ denotes the IAW of LDI, $v_{te}$ is the electrons thermal velocity and $c_s$ is phase velocity of IAW, then we can obtain 
\begin{equation}
\label{Eq:Delta_k}
\Delta k=\frac{2}{3}\frac{1}{\lambda_{De}}\frac{c_s}{v_{te}}.
\end{equation}
In H plasmas, $c_s\simeq\sqrt{Z_iT_e/m_i}$, $Z_i, m_i$ are the charge and mass of H ions. However, in CH (1:1) plasmas, for $k_s\simeq 2k_{L1}$, the frequency $\omega_s$ and thus the phase velocity $c_s=\omega_s/k_s$ of the IAW can be solved by\cite{Feng_2016POP,Feng_2016PRE}
\begin{equation}
\label{Eq:Dispersion_CH}
\epsilon_L(\omega_s,k_s=2k_{L1})=1+\sum_j \frac{1}{(k_s\lambda_{Dj})^2}(1+\xi_jZ(\xi_j))=0,
\end{equation}
where 
$Z(\xi_j)=1/\sqrt{\pi}\int_{-\infty}^{+\infty}e^{-v^2}/(v-\xi_j)dv$ is the dispersion function,
and $\xi_j=\omega_s/(\sqrt{2}k_sv_{tj})$ is complex and $\omega_s=Re(\omega_s)+i\gamma$, $\gamma$ is the Landau damping; $v_{tj}=\sqrt{T_j/m_j}$, $\lambda_{Dj}=\sqrt{T_j/4\pi n_jZ_j^2e^2}$ is the thermal velocity and the Debye length of specie $j$ ($j$ represents electrons, H ions or C ions). And $T_j, n_j, m_j, Z_j$ are the temperature, density, mass, and charge number of specie $j$, respectively. In this letter, $T_i/T_e=1/3$, the Landau damping of the slow mode is much lower than the fast mode, so the slow IAW mode will be excited in the case of CH plasmas.

\begin{figure}[!tp]
	\includegraphics[width=1\columnwidth]{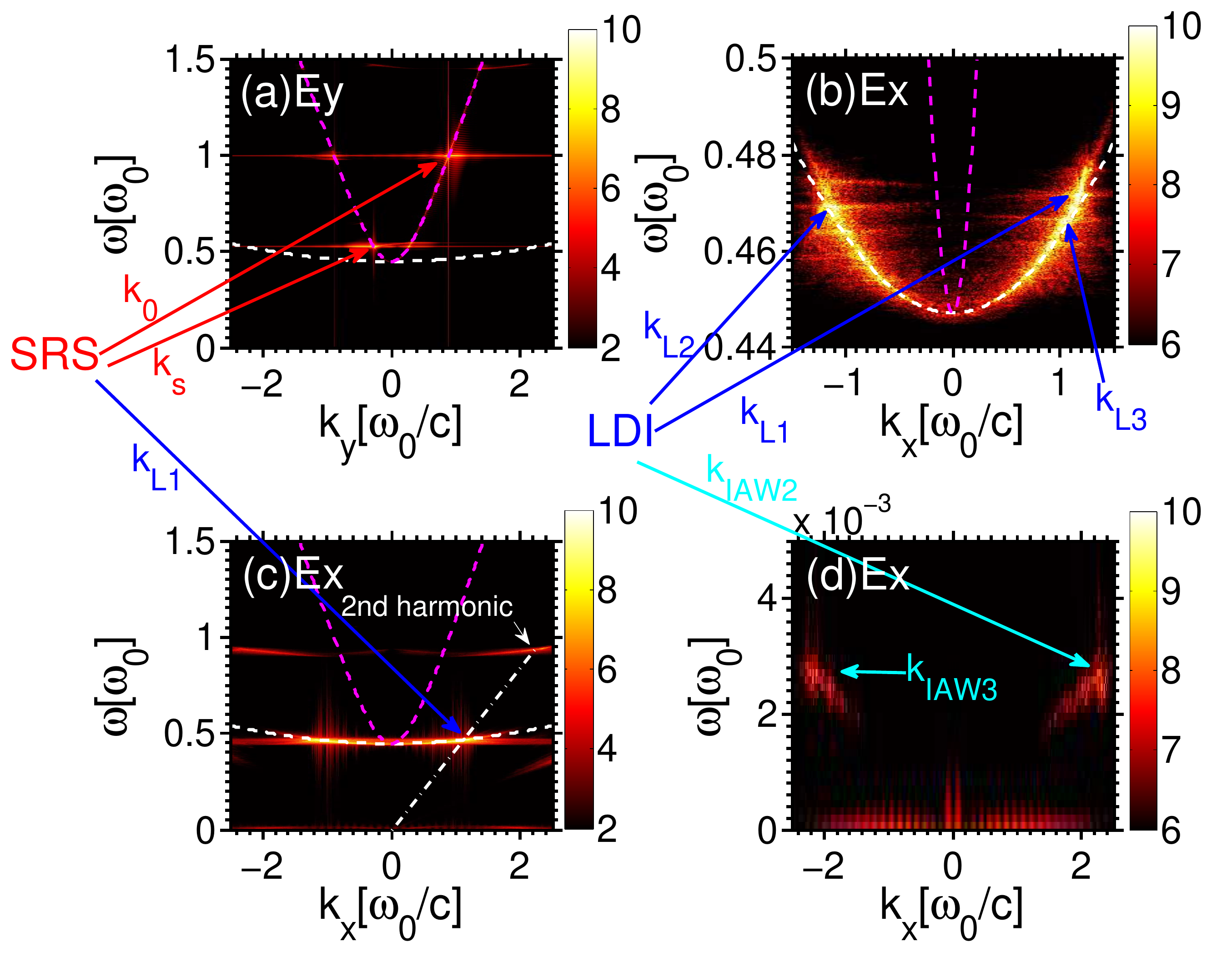}
	
	\caption{\label{Fig:Dispersion}(Color online) The dispersion relation of (a) the electromagnetic waves and (b)-(d) electrostatic waves in the case of CH plasmas.}
\end{figure}

Figure \ref{Fig:Dispersion} shows clearly the process of BSRS, LDI and multi-stage LDI (called LDI cascade). The corresponding wave numbers and frequencies of the LW and IAW generated by BSRS and LDI are listed in Table \ref{table1}. In the condition of $T_i/T_e=1/3$ in CH plasmas, $k_{IAW2}\simeq2k_{L1}=0.36\lambda_{De}^{-1}$, through Eqs. (\ref{Eq:Delta_k}) and (\ref{Eq:Dispersion_CH}), $c_s/v_{te}\simeq0.1558$, thus $\Delta k=0.0665\omega_0/c$. In our simulation, the value of $\Delta k$ is: $\Delta k=|k_{L1}|-|k_{L2}|=|k_{L2}|-|k_{L3}|=0.063\omega_0/c$. The simulation results are consistent to the theoretical calculations. The electrons temperature is $T_e=2.5keV$ and the electrons density is as large as $n_e=0.2n_c$, thus the BSRS rescatter \cite{Winjum_2013PRL} is excluded. We have checked the BSRS and stimulated Brillouin scattering (SBS) \cite{Merklein_2015NC,Bahl_2011NC,Shin_2013NC,Ballmann_2015SR} reflectivity (not shown here), the reflectivity of SBS is much lower than that of BSRS. So the SBS effect can be neglected.  These results verify that the LDI and LDI cascade occur and will dominate the saturation of BSRS. On the other hand, the continuous LW spectrum is due to the LW collapse\cite{Russell_1999POP}. However, we can distinguish several-order LWs (LW1, LW2, LW3, LW4, LW5, LW6) generated by the LDI cascade very clearly. Thus, the LDI cascade and Langmuir collapse will be the dominant saturation mechanism of BSRS.
\begin{table}
	\caption{\label{table1} The wave numbers and frequencies of LW and IAW generated by BSRS and LDI in CH plasmas. (The sign of the wave numbers represents the direction of the wave vectors. The wave numbers $k$ are normalized to $\omega_0/c$, the frequencies $\omega$ are normalized to $\omega_0$.)}
	\begin{ruledtabular}
		
		\begin{tabular}{c|c|c|c}
			\hline
			\bf{BSRS}& \bf{[$\bf{k_0}$, $\bf{\omega_0}$]}& \bf{[$\bf{k_s}$, $\bf{\omega_s}$]}&\bf{[$\bf{k_{L1}}$, $\bf{\omega_{L1}}$]}\\
			\hline
			{Theory} & [0.894, 1]&[-0.285, 0.531]&[1.180, 0.469]\\
			\hline
		{Simulation} &[0.888, 1]&[-0.283, 0.528] &[1.177, 0.472]\\
			\hline
			\hline
					
						\bf{1st LDI}& \bf{[$\bf{k_{L1}}$, $\bf{\omega_{L1}}$]}& \bf{[$\bf{k_{L2}}$, $\bf{\omega_{L2}}$]}&\bf{[$\bf{k_{IAW2}}$, $\bf{\omega_{IAW2}}$]}\\
						\hline
					{Theory} & [1.180, 0.469]&[-1.113, 0.4671]&[2.293, 0.0025]\\
						\hline
					{Simulation} &[1.177, 0.472]&[-1.114, 0.4695] &[2.298, 0.0026]\\
						\hline					
					\hline
					
					\bf{2nd LDI}& \bf{[$\bf{k_{L2}}$, $\bf{\omega_{L2}}$]}& \bf{[$\bf{k_{L3}}$, $\bf{\omega_{L3}}$]}&\bf{[$\bf{k_{IAW3}}$, $\bf{\omega_{IAW3}}$]}\\
					\hline
				{Theory} & [-1.113, 0.4671]&[1.047, 4.648]&[-2.160, 0.0024]\\
					\hline
				{Simulation} &[-1.114, 0.4695]&[1.051, 0.4666] &[-2.172 0.0025]\\
					\hline

		\end{tabular}
		
	\end{ruledtabular}
\end{table}

\begin{figure}[!tp]
	\includegraphics[width=0.8\columnwidth]{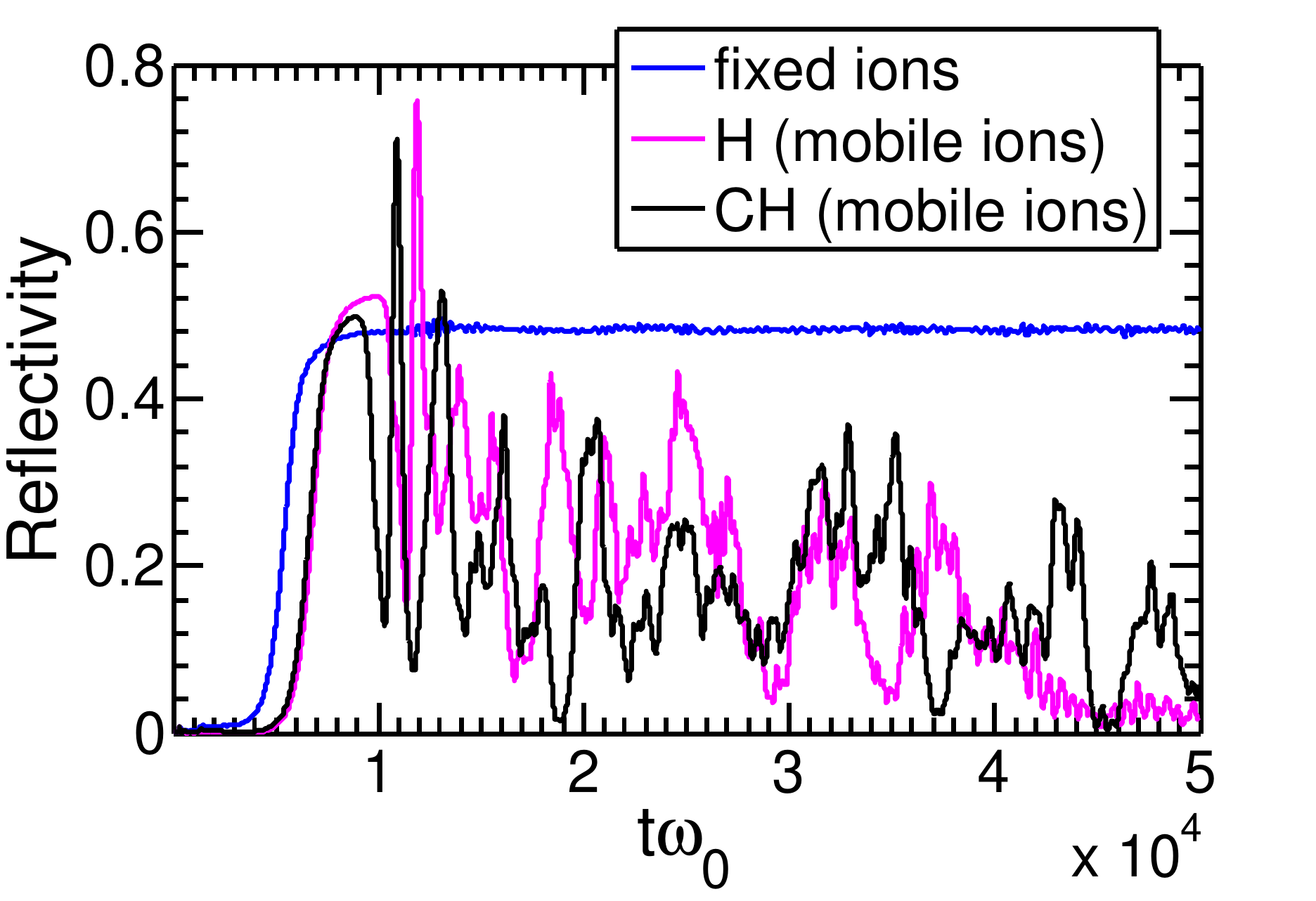}
	
	\caption{\label{Fig:Reflectivity}(Color online) The reflectivity of BSRS in the cases of fixed background ions, H mobile ions and CH mobile ions.}
\end{figure}

The BSRS reflectivities in different cases of species have been shown in Fig. \ref{Fig:Reflectivity}. We can find that the BSRS reflectivity in the case of fixed background ions is much larger than that in the cases of mobile ions. During the simulation time $[0, 5\times10^4\omega_0^{-1}]$, the average reflectivity of BSRS in the case of fixed background ions is 43.1\%, while the average BSRS reflectivity in the case of H mobile ions is 18.0\% and that in the case of CH mobile ions is 17.2\%. For the electrons fluctuation (produced by LWs) will make the mobile ions oscillate with a low frequency, the IAWs will be generated. The nature of this process is the Langmuir decay instability, which can transfer the energy of pump LW generated by BSRS to the IAW and the decay LW. And the decay LW is non-resonant with BSRS pump light and scattering light. At the same time, the energy of BSRS LW will be dissipated by Langmuir collapse. Thus, the BSRS LW energy will be reduced and the BSRS reflectivity will be reduced very much by LDI cascade and Langmuir collapse mechanism. As the Landau damping of slow IAW mode in CH plasmas ($\gamma_2/Re(\omega_s)=0.1212$) is a little lower than that of IAW in H plasmas ($\gamma_1/Re(\omega_s)=0.1884$), the LDI in CH plasmas will be stronger than that in H plasmas. Thus, the BSRS reflectivity in CH plasmas will be a little weaker than that in H plasmas.

\begin{figure}[!tp]
	\includegraphics[width=0.9\columnwidth]{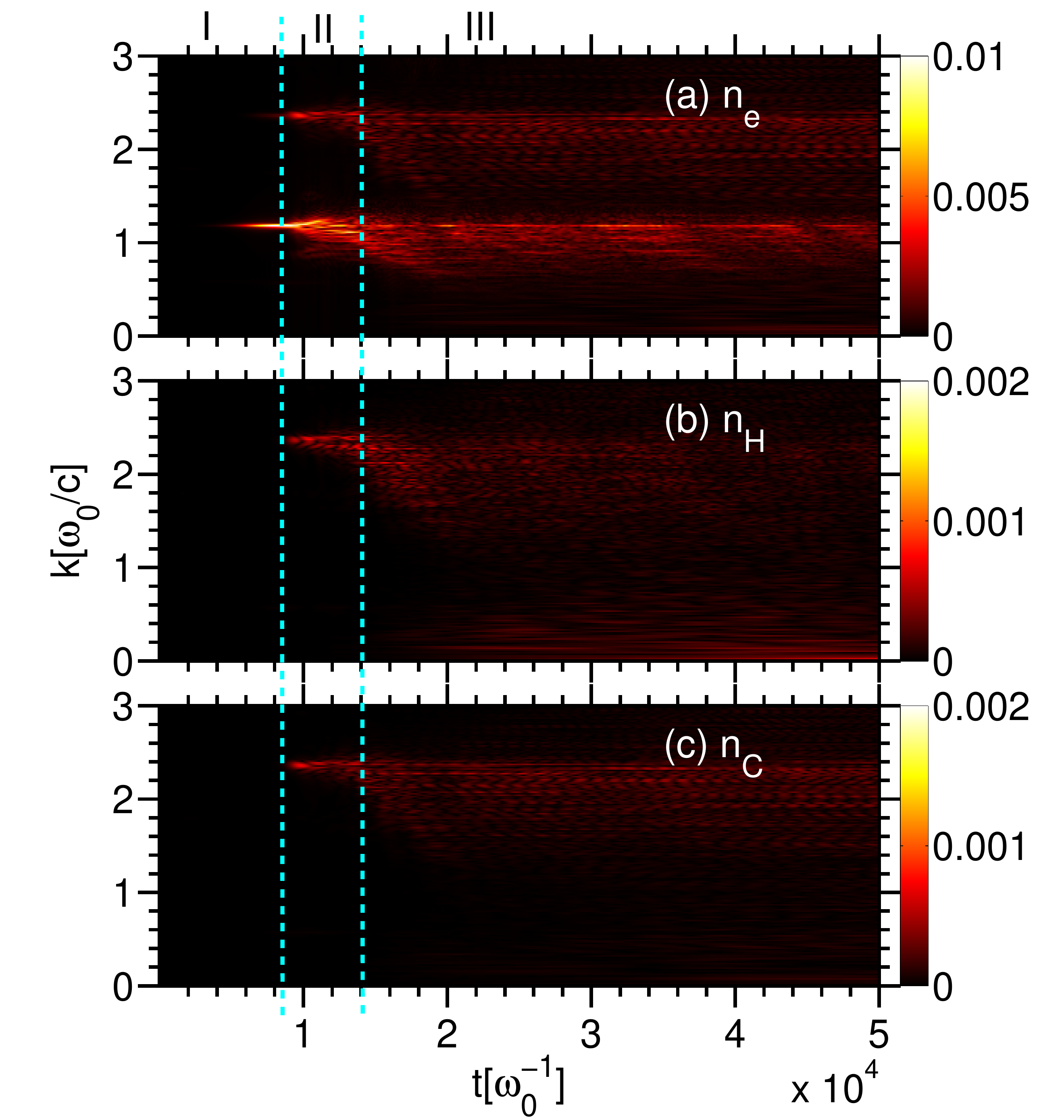}
	
	\caption{\label{Fig:k_t}(Color online) The wave-number time evolution of (a) electrons density fluctuation, (b) H ions density fluctuation and (c) C ions density fluctuation in the case of CH plasmas.}
\end{figure}

\begin{figure*}
	\begin{minipage}{0.8\textwidth} 
		\includegraphics[width=\columnwidth]{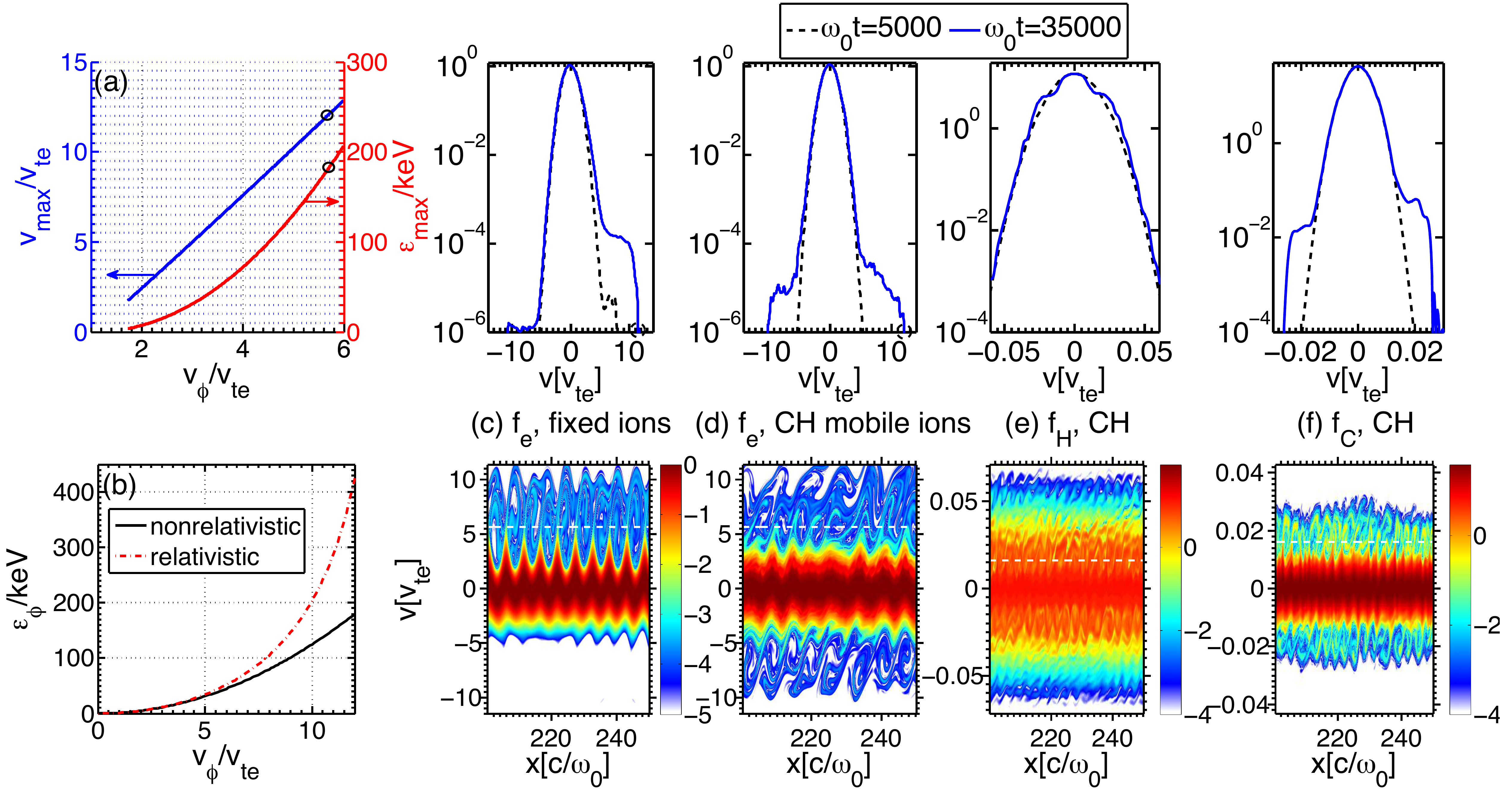}
	\end{minipage}%
	\begin{minipage}{0.2\textwidth}
		\caption{\label{Fig:PhasePicture}(Color online) (a) The wave-breaking maximum electrons-velocity and energy estimation in warm-nonrelativistic condition. (b) The energy corresponding to the LW phase velocity in the condition of nonrelativistic and relativistic condition. The space-average distributions and phase pictures of (c) electrons in the case of fixed background ions and (d) electrons, (e) H ions, (f) C ions in the case of CH mobile ions.}
	\end{minipage}
\end{figure*}

To clarify the LDI cascade process, the wave-number spectrum evolution with time is shown in Fig. \ref{Fig:k_t}. The time can be divided into three region approximately: (1) Region I, BSRS region, $\omega_0t\in[0,8300]$ and BSRS develops from $\omega_0t\simeq4500$; (2) Region II, BSRS+LDI region, $\omega_0t\in[8300,14000]$; (3) Region III, BSRS+LDI cascade region, $\omega_0t\in[14000, 50000]$. In Region I, the electrons density fluctuation spectrum (corresponding to the LW spectrum) shows a single-wave-number spectrum with time. The wave number is $k_{L1}=1.174\omega_0/c$ corresponding to LW generated by BSRS and keep constant. This result is consistent to the BSRS reflectivity as shown in Fig. \ref{Fig:Reflectivity} and interprets that the BSRS develops without LDI until $\omega_0t\simeq8300$. In Region II, the first LDI has developed. The LW wave-number spectrum (Fig. \ref{Fig:k_t}(a)) starts to demonstrate a broadening towards small wave numbers, due to the secondary LW produced in the first LDI process. Correspondingly, the wave number of the IAW (Figs. \ref{Fig:k_t}(b) and \ref{Fig:k_t}(c)) is $k_{IAW2}=2.348\simeq2k_{L1}$ and is generated by the first LDI. When LDI occurs, the BSRS reflectivity will decrease abruptly and burst after $\omega_0t\simeq8300$ (Fig. \ref{Fig:Reflectivity}). In Region III, after $\omega_0t\simeq1.4\times10^{4}$, the wave-number spectra of IAWs and LWs will display the further broadening towards small wave numbers, interpreted as the signature of the LWs and IAWs generated in the LDI cascade and Langmuir collapse. The LDI cascade and Langmuir collapse will result in a further decrease of the BSRS reflectivity after $\omega_0t\simeq1.4\times10^4$ (Fig. \ref{Fig:Reflectivity}). Note that Fig. \ref{Fig:k_t}(a) also shows the evolution of a wave-number spectra of the second harmonic corresponding to the fundamental LWs, which is related to the second harmonic shown in Fig. \ref{Fig:Dispersion}(c). However, the amplitude of the electric field of the second harmonic is about 3.5\% of that of the fundamental LWs. So the harmonic energy is much lower than the fundamental LWs energy and the energy loss from the harmonics generation can be neglected.

Figure \ref{Fig:PhasePicture} gives a snapshot of hot electrons generated from the electrons trapping in the cases of fixed background ions (Fig. \ref{Fig:PhasePicture}(c)) and CH plasmas (Fig. \ref{Fig:PhasePicture}(d)). The maximum velocity of the hot electrons from trapping is $v_{max}=v_{\phi}+v_{tr}=v_\phi+2\sqrt{e\phi_{max}/m_e}$, where $v_\phi$ is the phase velocity of the LW and $\phi_{max}=E_{max}/k$ is the maximum electric potential of LW. In the warm-nonrelativistic condition, the maximum wave-breaking amplitude is given by\cite{Coffey_1971POF, Mori_1990PS}
\begin{equation}
E_{max}=\frac{m_ev_\phi\omega_{pe}}{e}\sqrt{1+2\beta^{1/2}-\frac{8}{3}\beta^{1/4}-\frac{1}{3}\beta},
\end{equation}
where $\beta=3v_{te}^2/v_\phi^2$. And the wave number of the LW can be calculated from the LW dispersion relation:
\begin{equation}
k=\frac{1}{\lambda_{De}}\sqrt{\frac{v_{te}^2}{v_\phi^2-3v_{te}^2}}=\frac{1}{\lambda_{De}}\sqrt{\frac{\beta}{3-3\beta}}.
\end{equation}
Therefore, the maximum velocity of the hot electrons $v_{max}$ can be estimated from the wave-breaking amplitude $E_{max}$. Then, the maximum energy of hot electrons is $\varepsilon_{max}=\frac{1}{2}m_ev_{max}^2$ in the nonrelativistic condition and $\varepsilon_{max}'=(\gamma_{v_{max}}-1)m_ec^2$, $\gamma_v=\frac{1}{\sqrt{1-(v/c)^2}}$ in the relativistic condition. The relation of the maximum velocity $v_{max}$, the maximum energy $\varepsilon_{max}$ in the nonrelativistic condition with the phase velocity $v_\phi$ is shown in Fig. \ref{Fig:PhasePicture}(a). In our simulation, $v_\phi/v_{te}=5.7$, thus, the maximum velocity is $v_{max}/v_{te}\simeq12$, which is consistent to the maximum hot-electrons velocity as shown in Fig. \ref{Fig:PhasePicture}(c) and Fig. \ref{Fig:PhasePicture}(d). And the corresponding maximum energy is $\varepsilon_{max}\simeq180keV$ in nonrelativistic condition and $\varepsilon_{max}'\simeq427keV$ in relativistic condition. The energy at the LW phase velocity calculated by the relativistic and nonrelativistic condition is shown in Fig. \ref{Fig:PhasePicture}(b). However, the trapping electrons mainly gather around the LW phase velocity ($v_\phi/v_{te}=5.7$), and the number of superthermal electrons at the end of the distribution is very small. So the relativistic effect of the electrons around the LW phase velocity is not obvious here. Figs. \ref{Fig:PhasePicture}(c)-\ref{Fig:PhasePicture}(f) give the electrons distributions and ions distributions in the typical cases of fixed background ions and CH mobile ions. When the ions are mobile, the LDI cascade will occur, which can be observed from the signature of the ions trapped by the LDI IAW2 (positive phase velocity, denoted as $+$) and IAW3 (negative phase velocity, denoted as $-$) in Figs. \ref{Fig:PhasePicture}(e) and \ref{Fig:PhasePicture}(f). And also, the electrons will be trapped by the 
BSRS LW1 ($+$) and LDI LW2 ($-$), LW3 ($+$) as shown in Fig. \ref{Fig:PhasePicture}(d) (only the primary BSRS LW phase velocity is marked). As the LW energy of BSRS will transfer to the decay LW and IAW through LDI cascade process, at the same time, the energy of BSRS LW will be dissipated through Langmuir collapse, the BSRS LW energy will decrease and the decay LW energy is also lower than the BSRS LW energy. As a result, although the maximum velocity of the superthermal electrons is nearly as the same as the case of fixed background ions, the proportion of the hot electrons in the case of BSRS with LDI cascade (Fig. \ref{Fig:PhasePicture}(d)) is much lower than the BSRS without LDI (Fig. \ref{Fig:PhasePicture}(c)). In CH plasmas, when $T_i/T_e=1/3$, the Landau damping of the slow IAW mode ($\gamma_s/Re(\omega_s)=0.1212$) is much lower than that of the fast IAW mode($\gamma_f/Re(\omega_f)=0.691$)\cite{Feng_2016POP}. Thus, the slow IAW will be excited in LDI. These results demonstrate that LDI cascade is an effective mechanism to suppress the generation of hot electrons. 

In conclusion, we have demonstrated an effective mechanism to suppress BSRS and the hot electrons generation by Langmuir decay instability cascade and Langmuir collapse. For the cases of mobile ions is much closer to the realistic condition in NIF, the BSRS with LDI cascade and Langmuir collapse should be considered. Furthermore, in addition to changing plasmas parameter, laser parameter, we can control the LDI level, thus BSRS level, through changing ion species and ions ratio. This topic we believe is of universal interest to the field of laser particle interaction.

We are pleased to acknowledge useful discussions with L. Hao, T. W. Huang, K. Q. Pan and B. Qiao. This research was supported by the National Natural Science Foundation of China (Grant Nos. 11575035, 11475030 and 11435011) and National Basic Research
Program of China (Grant No. 2013CB834101).

\bibliography{LDI}

\end{document}